\documentclass[a4paper,fleqn,usenatbib]{mnras}

\usepackage[T1]{fontenc}
\usepackage{ae,aecompl}

\usepackage{graphicx}	
\usepackage{amsmath}	
\usepackage{amssymb}	

\newcommand{\rxte}{\textit{RXTE}}
\newcommand{\src}{\mbox{2A 1822-371}}
\newcommand{\ergs}{\rm\,erg\,s^{-1}}
\newcommand{\mdot}{\rm\,M_{\odot}\,yr^{-1}}
\newcommand{\msun}{\rm \,M_{\odot}\,}

\title{The X-ray Pulsar 2A 1822-371 as a Super Eddington source}
\author[A. B. Nielsen et al.]{
Ann-Sofie Bak Nielsen,$^{1}$\thanks{E-mail: nielsen@strw.leidenuniv.nl}
Alessandro Patruno,$^{1,2}$
Caroline D'Angelo,$^{1}$
\\
$^{1}$ Leiden Observatory, University of Leiden, Niels Bohrweg 2, 2333 CA, Leiden\\
$^{2}$ ASTRON, the Netherlands Institute for Radio Astronomy, Postbus 2, 7900 AA, Dwingeloo, The Netherlands}

\pubyear{2017}

\begin{document}
\label{firstpage}
\pagerange{\pageref{firstpage}--\pageref{lastpage}}
\maketitle

\begin{abstract}
The low mass X-ray binary 2A 1822-371 is an eclipsing system with an
accretion disc corona and with an orbital period of 5.57 hr. The
primary is an 0.59 s X-ray pulsar with a proposed strong magnetic
field of $10^{10}-10^{12}\rm\,G$. In this paper we study the spin
evolution of the pulsar and constrain the geometry of the system. We
find that, contrary to previous claims, a thick corona is not
required, and that the system characteristics could be best explained
by a thin accretion outflow due to a super-Eddington mass transfer
rate and a geometrically thick inner accretion flow.  The orbital,
spectral and timing observations can be reconciled in this scenario
under the assumption that the mass transfer proceeds on a thermal
timescale which would make \src\, a mildly super-Eddington source
viewed at high inclination angles.  The timing analysis on 13 years of
\textit{RXTE} data show a remarkably stable spin-up that implies
that \src\, might quickly turn into a millisecond pulsar in the next
few thousand years. 
\end{abstract}



\section{Introduction}
2A 1822-371 is a persistent eclipsing low mass X-ray binary (LMXB)
with a 0.59 s accreting X-ray pulsar
\citep{Jonker_2001ApJ...553L..43J}. The neutron star primary accretes
material from a 0.62M$_\odot$ Roche lobe filling main sequence star
\citep{Harlaftis_1997MNRAS.285..673H}, and the system has a binary
orbit of 5.57 hr
\citep{Jonker_2001ApJ...553L..43J,Hellier_1990MNRAS.244P..39H,Parmar_2000A&A...356..175P}. \citet{White_1981ApJ...247..994W}
showed that the partial eclipses of the system are best explained by
the presence of an accretion disk corona (ADC). The eclipses are
clearly seen because the system is being viewed almost edge on at an
inclination angle of 81-84$^\circ$ \citep{Heinz_2001MNRAS.320..249H,
  Ji_2011ApJ...729..102J, Jonker_2003MNRAS.339..663J}, which was
found through modelling of the light curve
\citep{Heinz_2001MNRAS.320..249H}. As shown by \citet{Heinz_2001MNRAS.320..249H} the eclipse of 2A 1822-371 is a narrow peak in the light curve, however, most of the light curve, about 80 \% of the orbit, is obscured. \citet{White_1982ApJ...257..318W}
suggested that the ADC is formed by evaporated material in the inner
accretion disc due to radiation pressure from the X-rays produced by
the neutron star.  The accretion disc is thought to be optically thick
and the ADC appears so extended that it is not completely blocked by
the companion star. Indeed the companion seems to eclipse about 50\%
of the total light emitted~\citep{Somero2012A&A...539A.111S,
  Mason_1982ApJ...262..253M}.
The magnetic field of 2A 1822-371 was inferred twice from the presence
of cyclotron resonance scattering features
(crsf). \citet{Sasano_2014PASJ...66...35S} reported results
obtained with \textit{Suzaku} and suggested a crsf at 33 keV
which would correspond to a magnetic field of B$\sim$2.8$\times$10$^{12}$G.
This, however,
was in disagreement with the later findings of 
\citet{Iaria_2015A&A...577A..63I} who interpreted
\textit{XMM-Newton} spectral data as showing a crsf at around 0.7
keV (and an inferred magnetic field of B$\sim$8.8$\times$10$^{10}$G).

The intrinsic X-ray luminosity (L$_X$) of 2A 1822-371 is currently not well
constrained. The first source of uncertainty comes from the distance,
which is not well known although it was estimated to be around
2-2.5kpc based on modelling of infrared and optical
observations\citep{Mason_1982ApJ...262..253M}.

\cite{Mason_1982ApJ...262..253M} estimated the luminosity to be
$L_X\sim 1.1\times10^{35}\left(d/1kpc\right)$ which, for a distance of
about 2.5 kpc is ${\sim}10^{36}$ erg s$^{-1}$.  Since the
pulsar is seen edge-on, its optical to X-ray luminosity ratio is
$L_{\rm opt}/L_X \sim 15-65$. This value is very anomalous among
LMXBs, which have a typical ratio of the order of ${\sim}1000$.  The
binary also shows a very large orbital period derivative of
$\dot{P}_{orb}$=1.5-2.1$\times$10$^{-10}$ss$^{-1}$ (implying a very fast
orbital
expansion~\citealt{Iaria_2015A&A...577A..63I,Burderi_2010A&A...515A..44B,Jain_2010MNRAS.409..755J}),
and thus it has been suggested that the binary is undergoing a highly
non-conservative mass transfer, with the neutron star accreting at the
Eddignton-limit and the rest of the material expelled from the donor
star via radiation pressure (e.g.,
\citealt{Iaria_2015A&A...577A..63I}).  This also suggests the
possibility that 2A 1822-371 is an Eddington limited source
\citet{Jonker_2003MNRAS.339..663J} which would then be
compatible with one of the magnetic field estimates inferred from the
crsf (i.e., $B=8.8\times$10$^{10}$G).  Another peculiar phenomenon in
2A 1822-371 is the fast spin up of the system, which gives an
extremely short spin-up timescale of order 7000 yr
\citep{Jonker_2001ApJ...553L..43J}. When looking at the ensemble of
slow accreting pulsars in LMXBs, the short spin-up timescale of 2A
1822-371 is not unique. Indeed short timescales have previously been
observed in LMXB pulsars such as 4U 1626-67 and GX 1+4, both of which
show torque reversals
\citep{Jonker_2001ApJ...553L..43J,Bildsten_1997ApJS..113..367B}.
The accretion torque reversal is a still poorly understood phenomenon
that occurs in some accreting pulsars and that causes a switch from
a spin-up to a spin-down (and vice-versa). One possible interpretation
of torque reversals is a transition between a Keplerian and a sub-Keplerian
flow in the inner portion of the accretion disk (e.g., \citealt{Yi_1997ApJ...481L..51Y}).

The spin evolution of 2A 1822-371 can therefore be explained with two
different scenarios: either the system has started accreting very
recently (if the currently observed spin-up truly represents the
secular evolution of the neutron star spin) or, alternatively, what we
are observing is simply a short-term effect with the current spin up
that will be possibly balanced by an episode of spin-down in the next
future. The latter scenario would make \src\, similar to the other
high field LMXB pulsars like 4U 1626-67 and GX 1+4. In those two
systems, the phenomenon of torque reversals occurs on timescales of
years. Other pulsars, such as Her X-1, Cen X-3 and Vela X-1 (the
latter two are high mass X-ray binaries) show variations on shorter
timescales of days to a few years (see for example Fig. 6 in
\cite{Bildsten_1997ApJS..113..367B}). Since several other LMXB pulsars
show changes in their spin frequency derivative, we investigate here
whether long-term spin-up in 2A 1822-371 is really stable or if there
are underlying detectable fluctuations related to accretion torque variations.

A second problem is that, at least in the (low magnetic field)
accreting millisecond pulsars, measuring the spin frequency derivative
is sometimes complicated by the presence of timing noise in the X-ray
time of arrivals of pulsations \citep{Hartman_2008ApJ...675.1468H,
  Patruno_2009ApJ...698L..60P}. It has, however, been observed that at
least in some accreting millisecond pulsars, a large part of timing
noise is correlated to variations in flux
\citep{Patruno_2009ApJ...698L..60P, Patruno_2010ApJ...717.1253P, 2011ApJ...738L..14H}. One plausible interpretation of the flux-phase
correlation is that the hot spot is moving on the pulsar surface in
response to variations of the mass accretion rate.  In high field
accreting pulsars like 2A 1822-371 the presence of such correlation
has never been reported and it is currently unclear whether such
effects might be present in these systems too. For example, the
stronger magnetic field of the neutron star in 2A 1822-371 might
prevent a drift of the hot spot when the accretion rate varies.
However, strong timing noise has been observed in the accreting X-ray
pulsar Terzan 5 X-2 \citep{Patruno_Alpar_2012ApJ...752...33P}, which
is an 11 Hz accreting pulsar with a magnetic field of the order of
$10^9-10^{10}$ G, which is substantially stronger than the typical
field observed in accreting ms pulsars ($B\sim10^8\rm\,G$). Therefore
it might be possible that the same phenomenon is present in 2A
1822-371 and in this work we plan to investigate this.  There are
therefore two questions that need to be addressed for 2A 1822-371:
\begin{itemize}
  \item[1.] Is the previously measured spin frequency derivative the true one or
    its measurement is affected by the presence of timing noise?
  \item [2.] Does the (true) spin frequency derivative represent the long
    term spin evolution of the neutron star? 
\end{itemize}
In this paper we use archival data from the \textit{Rossi X-ray Timing
  Explorer} (\textit{RXTE}), collected over a baseline of 13 years, to
try to answer the aforementioned questions.

Furthermore, it has been suggested that the ADC forms an optically
thick region around the neutron star
\citep{Parmar_2000A&A...356..175P, Iaria_2001ApJ...557...24I,
  White_1982ApJ...257..318W}, with optical depth $\tau \sim 9-26$,
which implies that most of the light coming from the pulsar is heavily
scattered. However, at such large optical depths the coherent
pulsations cannot preserve their
coherence. \citet{Iaria_2013A&A...549A..33I} first noticed this
problem and suggested that the Comptonised component is produced in
the inner regions of the system which are never directly observed.
Then only a small fraction ($\sim1\%$) of the total light produced is
scattered along the line of sight of the observer with an optical
depth $\tau\sim0.01$. This geometry, although possible, requires some
fine tuning of the optical depth. Furthermore, to preserve the
coherence of the pulsations, the whole (optically thick)
Comptonization region needs to rotate with the neutron star.  A third
question that needs to be answered is therefore whether it is possible
to keep a simple geometry of the system, with an ADC, and still obtain
a spectrum compatible with $\tau{\sim}1$.

In section
\ref{sec: observation} we present the observations and the data
reduction procedure, in section \ref{sec:results} we show our results
on the timing analysis,
e.g. the spin evolution and flux-phase correlation, and in section
\ref{sec:Discussion} we discuss the implications of our finding and
we extend previous models for \src\,.

\section{X-Ray Observations} \label{sec: observation}
We have used data taken over a baseline of 13 years, between 28 June
1998 and 30 November 2011. All observations were taken with the
Proportional Counter Array (PCA) on board the \textit{RXTE}
\citep{Bradt_1993A&AS...97..355B,
  Jonker_2001ApJ...553L..43J}. \rxte\textbf{/}PCA consists of five
xenon/methane proportional counter units, which are sensitive in the
range of 2-60 keV \citep{Jahoda_2006ApJS..163..401J}.  We chose
science event files with a time resolution of 2$^{-16}$s
(Event\_16us), 2$^{-13}$s (Event\_125us) and 2$^{-20}$s (GoodXenon)
for the timing analysis whereas we selected Standard 2 data with 16 s
time resolution to construct the X-ray lightcurve.  The lightcurve is
reconstructed in the 2--16 keV energy range and the X-ray flux is first
averaged for each observation (ObsID) and then normalized in Crab
units (see Fig. \ref{fig:LightCurve_v1}). A detailed description of
this standard procedure can be found in
\citet{vanStraaten_2003ApJ...596.1155V}. The timing analysis is
performed by selecting the absolute energy channels 24 to 67 that
correspond approximately to an energy range of $\approx$9-23 keV. This
range was chosen because the pulsation have the highest signal to
noise ratio (S/N) as reported by \citet{Jonker_2001ApJ...553L..43J}.
The data was barycentered with the FTOOL \textit{faxbary} by using the
JPL DE405 Solar System coordinates and the most precise astrometric
position found by \textit{Chandra} observations, RA: 18:25:46.81 and
DEC: -37:06:18.5 \citep{Burderi_2010A&A...515A..44B} with an error of
0.6$"$ (90\% uncertainty circle of the \textit{Chandra}
X-ray absolute position).  The barycentered data was then epoch folded
in pulse profiles of 32 bins over either 1500 s, or the total length of
the data segment, which usually corresponds to 3000 s.  We then
cross-correlated each pulsation with a sinusoid at the spin frequency
of the pulsar and generated a set of time of arrivals (TOAs).  We
selected only pulsations \textbf{with} a S/N larger than 3.1$\sigma$,
defined as the ratio between the pulse amplitude and its 1 sigma
statistical error. The value of 3.1$\sigma$ is selected to account for
the number of trials, i.e., we expect less than 1 false pulse detections among
the 281 pulse profiles generated over the entire 13 year baseline.  We
looked for the presence of a 2$^{nd}$ harmonic which was detected
only is a small subset of data segments and thus we do not consider it any
further in the forthcoming timing analysis.

The ephemeris used in the epoch folding are composed by an initial pulse
frequency from \citealt{Jonker_2001ApJ...553L..43J} and
\citealt{Jain_2010MNRAS.409..755J} and an orbital solution from
\citealt{Iaria_2011A&A...534A..85I}. The orbital solution corresponds
to a Keplerian circular orbit with a constant orbital period
derivative. Since the data contains large data gaps we split it in 20
segments that could be potentially phase connected, meaning that our
time baseline for each different segment spans typically a few days,
see table \ref{tab:overview}.  With the S/N criterion discussed above,
we found significant pulsations in 15 out of 20 data segments. In this
work we have used TEMPO2 version 2012.6.1 and kept the orbital values
fixed throughout the analysis.

\begin{table*}
  \centering
  \caption{Overview of the data used throughout this paper. The data are taken with \textit{RXTE}, and in this data, segments 3, 9, 10, 13, 14, 15 represent data not analysed before in the literature \citet{Jain_2010MNRAS.409..755J, Somero2012A&A...539A.111S, Jonker_2001ApJ...553L..43J}. The errors reported in the parentheses correspond to $1\sigma$ statistical errors. The $\chi^2$ and dof given in the table is for the TEMPO2 spin frequency fit to the data. The 95\% upper confidence limit is only given for the data segments were there were not found any significant pulsations (significance limit was 3.3$\sigma$).}
  \label{tab:overview}
  \begin{tabular}{lccccccccr}
    \hline
    \# segment & ObsId & T$_{start}$ & T$_{end}$ & Upper limit ( \%) & PEPOCH & Spin frequency ($\nu_0$) & Spin period (P$_s$) & $\chi^2$ & dof\\
    \hline
    1 & 30060 & 50992.8210 & 50993.9805 & & 50993.1873 & 1.6860862(6) & 0.59308949(2) & 28.49 & 6 \\
    2 & 30060 & 51018.2434 & 51019.6464 & & 51018.9449 & 1.6860957(3) & 0.59308615(9) & 63.30 & 17 \\
    3 & 50048 & 52031.5426 & 52032.8178 & & 52032.1105 & 1.6866984(4) & 0.5928742(3) & 35.43 &  13\\
    4 & 50048 & 52091.4522 & 52101.4647 & & 52096.4049 & 1.6867365(2) & 0.59286083(7) & 47.41 & 12\\
    5 & 70036 & 52435.4107 & 52435.7670 & & 52435.5174 & 1.686932(2)  & 0.59279212(5) & 25.54 & 8\\
    6 & 70037 & 52488.3701 & 52495.9115 & & 52491.9201 & 1.68697104(4) & 0.5927784(2) & 100.13 & 26\\
    7 & 70037 & 52503.3513 & 52503.9363 & & 52503.5256  & 1.686980(2) & 0.59277526(5) & 21.77 & 8\\
    8 & 70037 & 52519.1659 & 52519.5110 & & 52519.3091 & 1.686988(4) & 0.59277244(1) & 42.79 & 6\\
    9 & 70037 & 52547.3079 & 52547.8544 & & 52547.3944 & 1.6870063(10) & 0.59276601(4) & 26.31 & 9 \\
    10 & 70037 & 52608.1849 & 52608.3845 & & 52608.2072  & 1.686879(8)  & 0.59281075(3) & 11.22 & 2\\
    11 & 70037 & 52882.0214 & 52882.2226 & & 52882.3290 & 1.687240(4) & 0.59268391(2)  & 8.81 & 3\\
    12 & 70037 & 52883.1514 & 52885.1249 & & 52883.4896  & 1.6869901(8)  & 0.59277171(3) & 20.55 & 2\\
    13 & 80105 & 52896.2822 & 52896.4384 & & 52896.2556 & 1.687243(3)  & 0.59268286(9) & 11.12 & 4\\
    14 & 96344 & 55880.6677 & 55884.8487 & & 55882.5852 & 1.6891897(2) & 0.59199982(6) & 85.27 & 17\\
    15 & 96344 & 55888.5020 & 55895.6318 & & 55891.8696 & 1.68920448(8)  & 0.59199464(3) & 23.85 & 7\\
    16 & 60042 & 52138.7535 & 52141.8115 & 2.1 & - & - & - & - & -\\
    17 & 60042 & 52138.7535 & 52141.8115 & 2.1 & - & - & - & - & -\\
    18 & 50048 & 51975.7203 & 51976.1843 & 0.7 & - & - & - & - & -\\
    19 & 70037 & 52432.3887 & 52432.7906 & 0.6 & - & - & - & - & -\\
    20 & 70037 & 52724.7351 & 52724.7545 & 0.5 & - & - & - & - & -\\
    \hline
  \end{tabular}
\end{table*}

\section{Results}\label{sec:results}

Pulsations are found in the first 15 data segments described in table
\ref{tab:overview}. In the last 5 data segments the S/N found was
lower than the selected 3.1$\sigma$ for most of the pulsations, thus
leaving us with too few or no pulsations to perform the timing
analysis.  In table \ref{tab:overview} we provide the 95\% confidence
upper limit of the fractional amplitude of the pulsations in these 5
data segments. All these upper limits are consistent with
the fractional amplitude found in the first 15 data segments.

\subsection{Spin Evolution}\label{sec:LT_spin}

In two out of fifteen data segments the pulse time of arrivals of the
neutron star required a spin frequency derivative.  The $\chi^2$ and
degrees of freedom (dof) for the fits to all 15 data segments are
given in table \ref{tab:overview}. The errors on the spin frequency
were calculated by using standard $\chi^2$ minimization techniques and
by multiplying them by the square root of the reduced $\chi^2$. The
errors on the spin period is found from:
$\sigma_{P}$=$\sigma_{\nu}(\frac{1}{\nu})^2$.

\begin{figure}
  \centering
\includegraphics[width=1.1\linewidth]{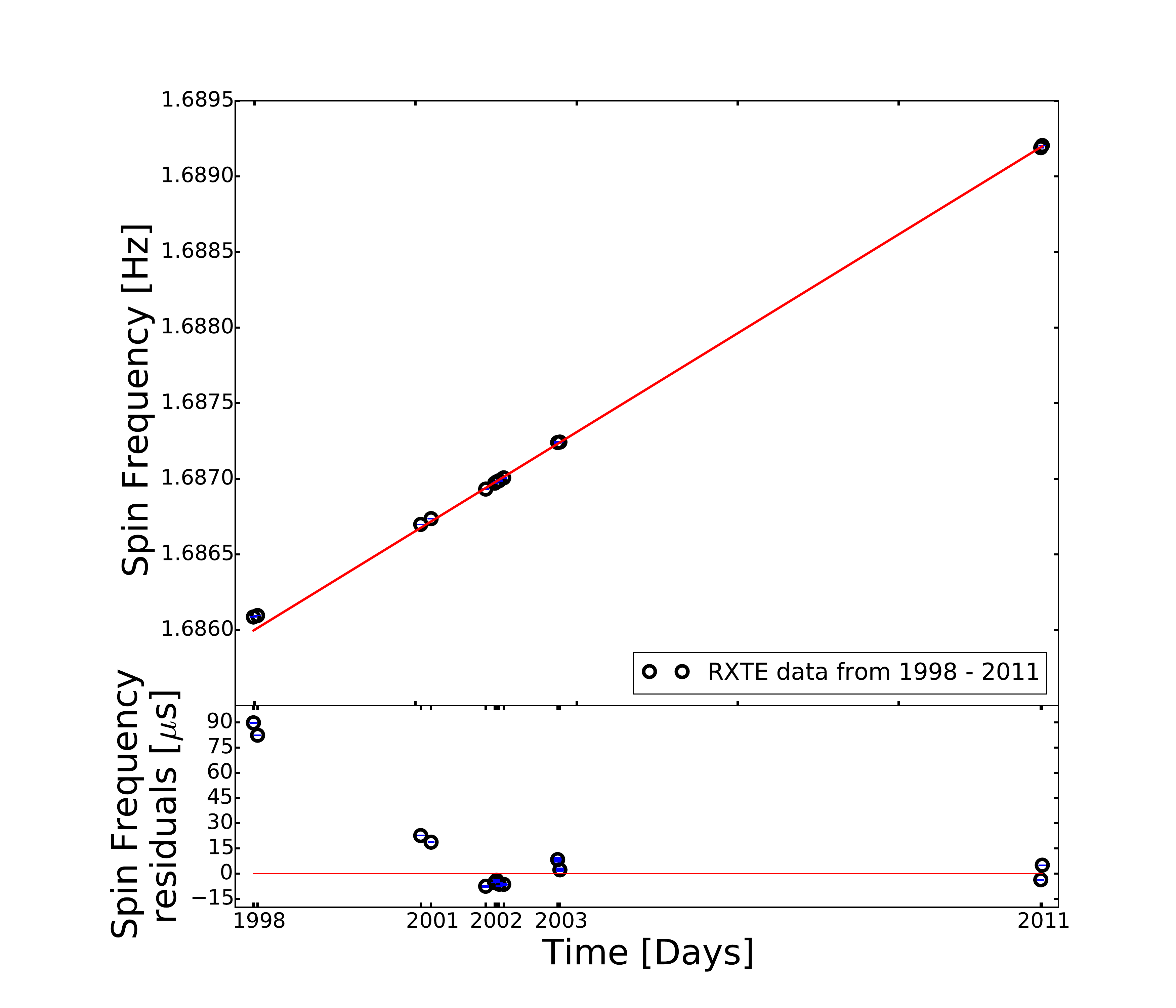}
  \caption{ The longterm spin up over a time period of 13 years. The
    black points show the data used in this paper and listed in table
    \ref{tab:overview}. The red line is a fit to the data from this
    paper and it gives the spin up of the pulsar of
    $\dot{\nu}$=7.57(6)$\times\,10^{-12}\rm\,Hz\,s^{-1}$. This figure is plotted
    with errors on all points, seen in blue. Most of the errors are,
    however, so small that they are not visible on the plot.  }
  \label{fig:period}
\end{figure}

The collection of all spin frequencies (see Table~\ref{tab:overview})
is then fitted with a linear function to determine the long-term spin
frequency derivative over the entire baseline of the observations.
The fit gives a reduced $\chi^2$ of 1909 for
11 dof. This indicates that the fit is not
  statistically acceptable (p-value$<0.05\%$) and the bad fit is caused
by several points that are clearly off the linear relation (see lower panel of
Fig. \ref{fig:period}).

The best-fit linear slope that we find corresponds to a spin up of
$\dot{\nu}$=7.6(8)$\times\,10^{-12}\rm\,Hz\,s^{-1}$ for the period
ranging from June 1998 to November 2011, comparable to what has recently been found by \cite{Chou_2016arXiv160804190C}.

The bad fit to the spin frequencies shows that, although there is no
evidence for a change in sign of the spin frequency derivative, some
fluctuations are present in the data. We thus explored to what extent the magnitude of the accretion torque vary with time and whether
\textit{small/short-term} torque reversals are present in the data. 
First, in the data segments 4 and 6, it was possible to
phase-connect the pulsations and we thus have a direct measure of the
spin frequency derivative. Such spin frequency derivative is measured
for a time interval of 10 and 8 days (for segment 4 and 6,
respectively). The two spin
up values found are $\dot{\nu}$=6.7(4)$\times\,10^{-12}\rm\,Hz\,s^{-1}$ for segment
4 and $\dot{\nu}$=8.2(5)$\times\,10^{-12}\rm\,Hz\,s^{-1}$ for segment 6 which are both
within $1\sigma$ from the long-term linear trend seen in the 13-year long
baseline.

\begin{figure}
\centering
\includegraphics[width=1.0\linewidth]{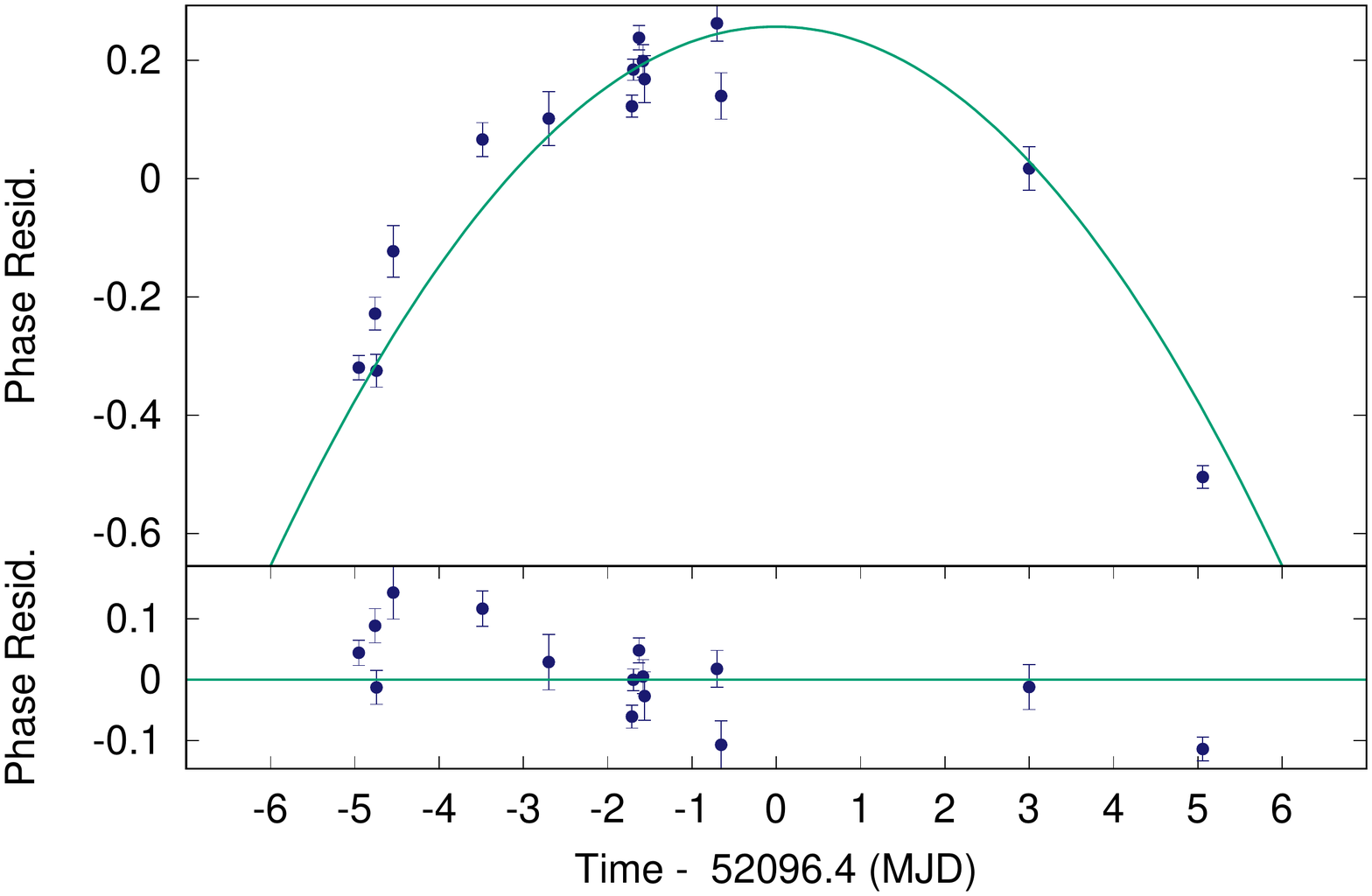}
\caption{The pulse phase residuals before a spin frequency derivative is fitted for (top panel) and after (lower panel). The parabola is a clear evidence of the presence of a spin frequency derivative.}
\label{fig:july01_parabola}
\end{figure}

The pulse phase residuals with respect to a constant spin frequency
model can be seen for the data segment 4 in the top panel of
Fig. \ref{fig:july01_parabola}. In the lower panel of the same figure
we show the pulse phase residuals with respect to a spin frequency
derivative model. The parabolic trend seen in the top panel of
Fig. \ref{fig:july01_parabola} is a clear signature of the presence of
a frequency derivative.

\subsection{Spin period derivative vs. time}\label{sec:P_dot vs. time}

\begin{figure}
\centering
\includegraphics[width=1.1\linewidth]{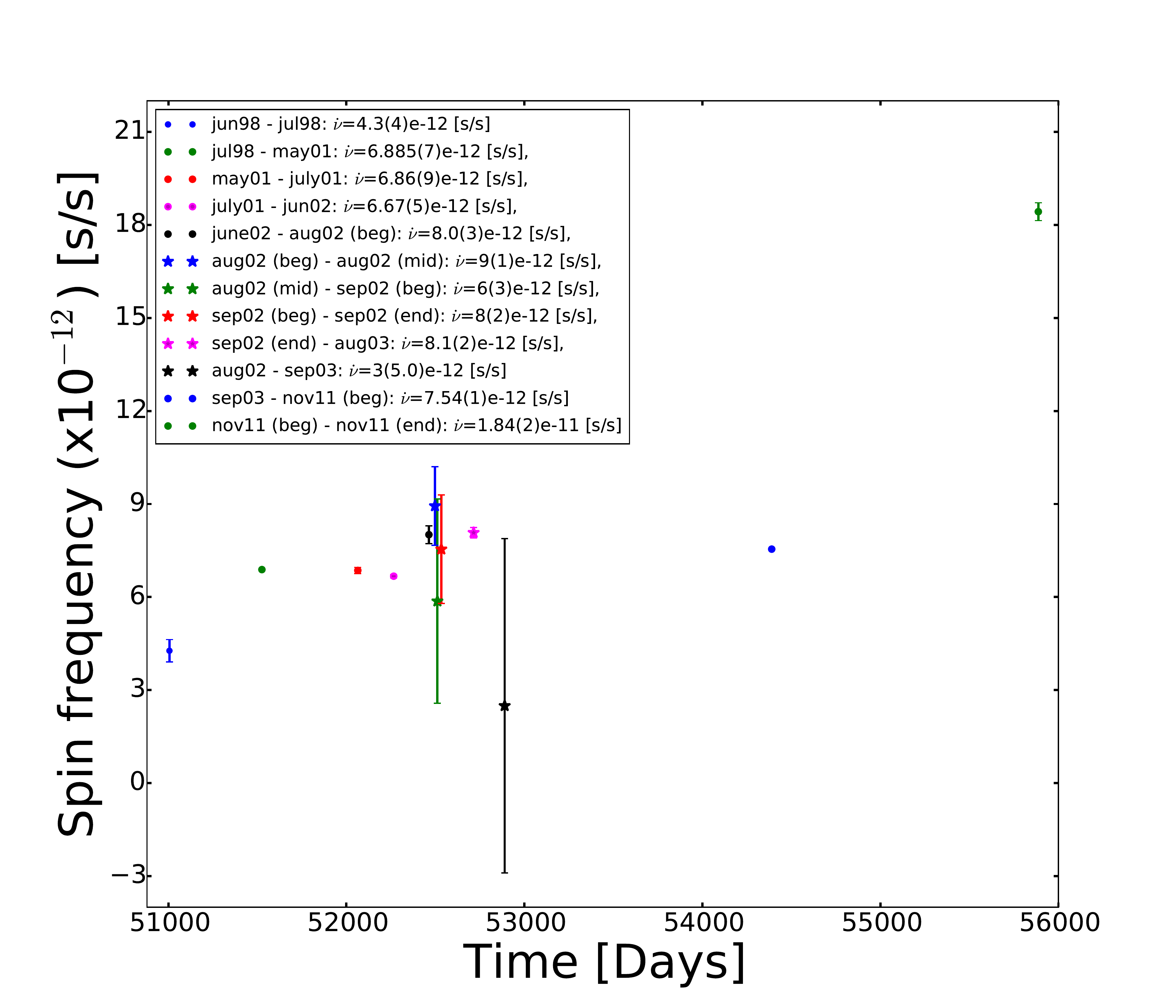}
\caption{The spin period derivative change between the individual observations from Fig. \ref{fig:period}. It is clearly seen that most of the observations coincide, and there is a close to constant development in the spin period derivative. However, there are a few points that are off, e.g. in 2003 and 2011.}
\label{fig:Ps_T}
\end{figure}

To estimate the magnitude of the spin fluctuations we calculated the
spin period derivative between each consecutive data point, i.e., we
determined the slope between each pair of points, and plotted them
versus time, see Fig. \ref{fig:Ps_T}.  The errors on the spin period
derivatives are found by taking the maximum and minium slope between
the spin periods in Fig. \ref{fig:period}.  It is clear that most of
the points in Fig. \ref{fig:Ps_T} are roughly consistent with a single
constant spin period derivative with small variation of less than a
factor 2 over the whole baseline with the exception of an outlier 
in the last data segment taken in 2011. In that case the
spin frequency derivative is consistent with having increased by a
factor 3 during the last month of the observations.

\begin{figure}
\centering
\includegraphics[width=1.1\linewidth]{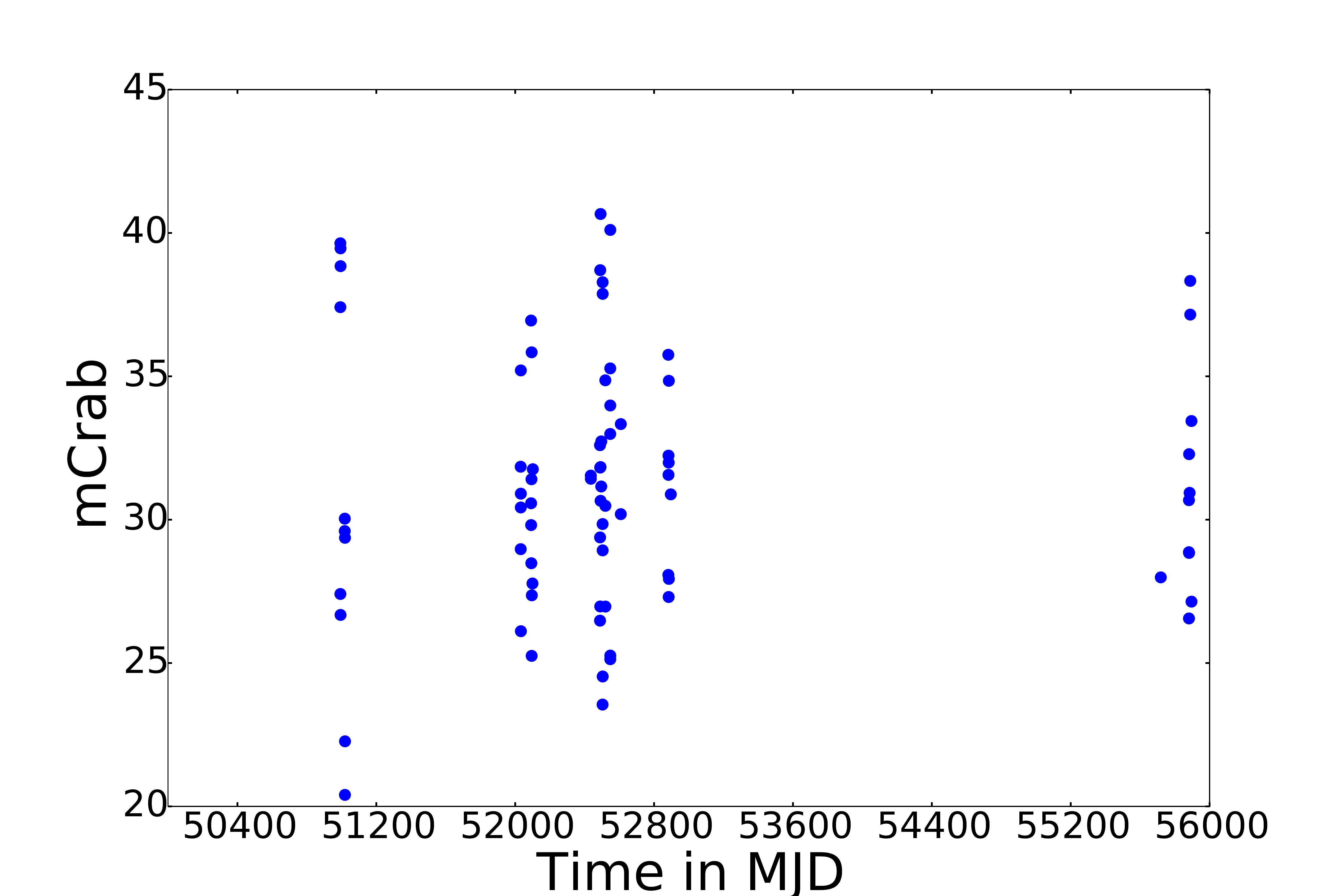}
\caption{2--16 keV X-ray lightcurve of 2A 1822-371. Each data point
  represents an ObsID average flux. The overall intensity of the light
  curve is fairly constant over the span of the observations used in
  this paper, with variations of less than a factor 2 in 13 years of observations.}
\label{fig:LightCurve_v1}
\end{figure}

The 2--16 keV X-ray flux shows a variation of less than a
factor 2 during that same month (see Figure~\ref{fig:LightCurve_v1})
and no average variation when compared to the average X-ray luminosity
of the previous years.

\subsection{X-Ray Flux-Phase Correlation} \label{sec: flux-phase}

2A 1822-371 is a persistent source that shows little variability in
X-ray flux. In all our observations the X-ray flux varies by less than
a factor of 2 with respect to the average value. Therefore even if a
flux-phase correlation is present in 2A 1822-371 we expect little or
no variation in the X-ray pulse phases. Nonetheless we inspected the
data for the presence of such correlation, since it is the first time
that such a test is performed in such a high field accreting pulsar.

We follow the procedure outlined in
\citet{Patruno_2009ApJ...698L..60P,Patruno_2010ApJ...717.1253P} i.e.,
we minimize the $\chi^2$ of a linear fit to the X-ray flux vs. pulse
phase.  If there is a significant correlation then this might indicate
the existence of some mechanism that determines the pulse phase
variations in addition to genuine neutron star spin variations.  Such
mechanism for example can be the motion of the hot spot on the surface
of the pulsar
\citep{Patruno_2009ApJ...698L..60P,Patruno_2010ApJ...717.1253P}.

We fit the data with a linear correlation
\begin{equation}
\phi=a+b\,F_x
\end{equation}
 where $\phi$ is the pulse phase and $F_x$ the X-ray flux. 
If there is a correlation, $b$ should be significantly different than
zero. However, in all our 15 data segments, the $b$ coefficient is
consistent with zero within the statistical errors as can be seen in
table \ref{tab:Flux_Phase}.

\begin{table}
	\centering
	\caption{The A and B value for the Flux-Phase correlation fits for the individual data segments.}
	\label{tab:Flux_Phase}
	\begin{tabular}{lcr} 
		\hline
		\# segment & a & b \\
		\hline
		1 & -0.08(10) & 0.001(2) \\
		2 & 0.20(3) & -0.0087(7)\\
		3 & 0.2(1) & -0.007(4) \\
		4 & 0.81(6) & -0.03(2)\\
		5 & 0.08(7) & -0.003(2)\\
		6 & 0.004(40) & -0.0001(15)\\
		7 & 0.19(5) & -0.008(2)\\
		8 & -0.22(5) & 0.010(2) \\
		9 & 0.12(5) & -0.005(2)\\
		10 & -0.5(1) & 0.019(4)\\
		11 & -1.8(2) & 0.080(7)\\
		12 & 0.11(8) & -0.004(3)\\
		13 & 0.4(1) & -0.012(3)\\
		14 & -0.07(5) & 0.002(2)\\
		15 & -1.20(7) & 0.033(2)\\
		\hline
	\end{tabular}
\end{table}

\subsection{Fractional Amplitude vs. cycles}\label{sec:FA}

We tested whether there was a correlation between the fractional
amplitude of the pulsations and the eclipses with the purpose of testing
whether we can actually directly see the surface of the neutron star when there
is no eclipse.

\begin{figure}
\centering
\includegraphics[width=1.1\linewidth]{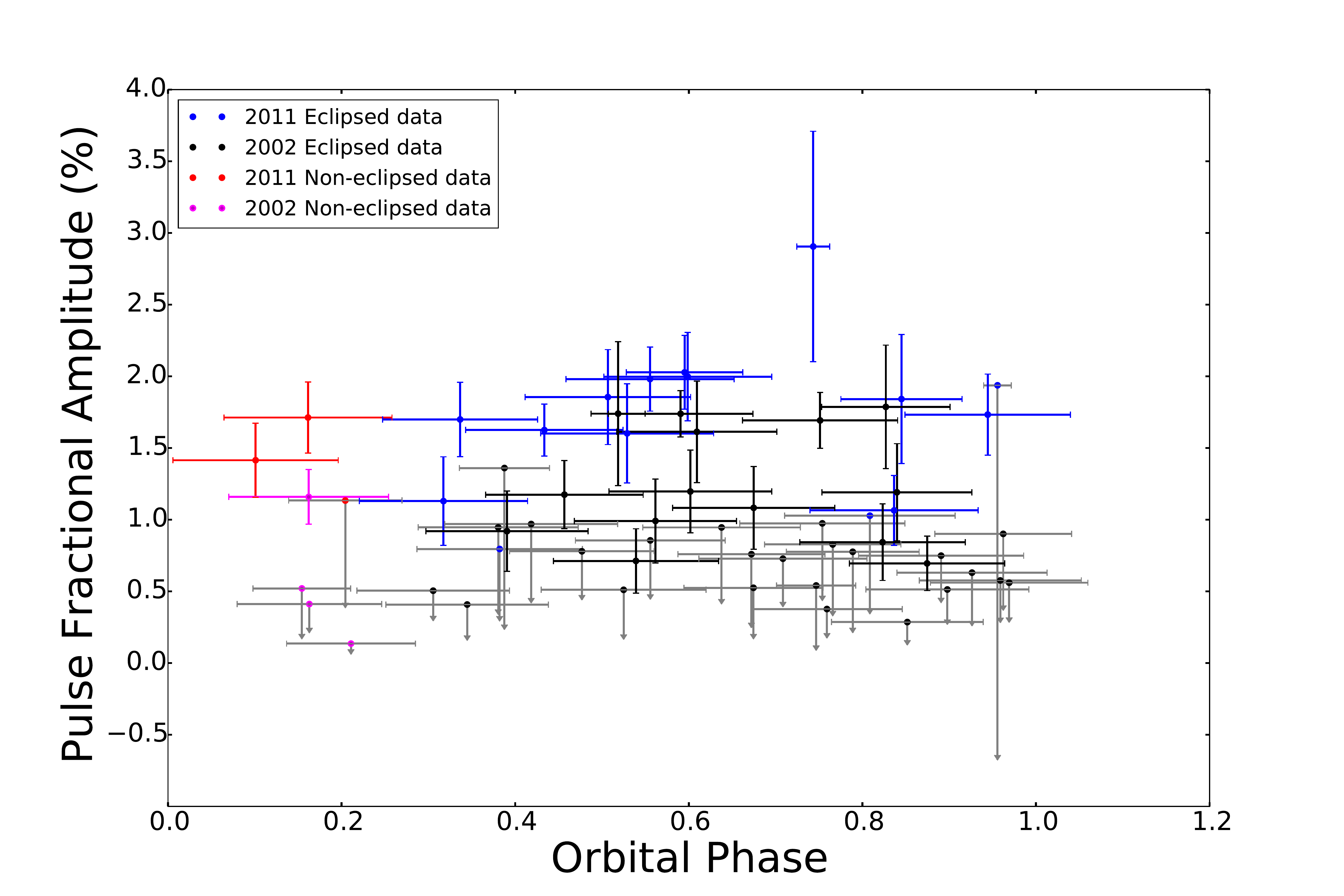}
\caption{The fractional amplitude of eclipsed data and the non-eclipsed data vs. the cycle of the pulsar. The Eclipse fractional amplitudes are seen in blue/green (2011 and 2002 data), and the non-eclipsed fractional amplitudes are seen in red/pink (2011 and 2002). The non-eclipsed data is assumed to be 0.05-0.25 cycle, and the rest of the cycle is assumed to be a part of the eclipse. The eclipse is defined with a zero point at T$_{ASC}$.}
\label{fig:FA_cycle}
\end{figure}

On Fig. \ref{fig:FA_cycle} we report the results of the data analyzed
from 2002 and 2011. The fractional amplitude definition is the same as
the one used in \citet{Patruno_2010ApJ...717.1253P}.  When pulsations
are not detected we provide 95\% confidence level upper limits. The
red and pink points on Fig. \ref{fig:FA_cycle} are the data from the
non-eclipsed part of the lightcurve, and the blue and black points are
from the eclipsed part of the orbit. We use T$_{\rm asc}$ as our
reference orbital phase zero (i.e., the beginning of the cycle) and we
thus expect that the non eclipsed data are between cycle 0.05-0.25. We
further use the fact that the rest of the cycle is partially
eclipsed. Note some other authors use a different definition of eclipse, for example
\cite{Heinz_2001MNRAS.320..249H}, define the eclipse to be only 
the portion of the lightcurve that shows a deep dip~\citep{Heinz_2001MNRAS.320..249H,
  Hellier_1990MNRAS.244P..39H}. In any case, Fig. \ref{fig:FA_cycle} shows that
there is no detectable difference between the fractional amplitudes for the
eclipsed and non eclipsed data in any orbital phase interval.

\section{Discussion}\label{sec:Discussion}
In this paper we have found a long-term spin frequency derivative of
$\dot{\nu}$=7.6(8)$\times10^{-12}\rm \,Hz\,s^{-1}$, and a short term
spin frequency derivative of $\dot{\nu}$=(6--8)$\times10^{-12}\rm
,Hz\,s^{-1}$, in two phase connected data segments. The long-term spin
up is consistent with that reported by
\cite{Chou_2016arXiv160804190C}, \cite{Iaria_2015A&A...577A..63I},
\cite{Jain_2010MNRAS.409..755J} and
\cite{Jonker_2001ApJ...553L..43J}. We find no evidence for torque
reversals and few variations in the accretion torque, limited to
fluctuations of less than a factor two with the exception of the last
data segment in 2011, where the spin up requires an increases by a
factor 3 with respect to the overall long-term spin frequency
derivative. There is no corresponding increase in the X-ray flux (in
the 2--16 keV band) at the time of the spin-up increase.  Finally, no
evidence for a phase-flux correlation and no strong variations in the X-ray
flux of the source are found.

Previous papers have suggested different models and parameters for
2A 1822-371. A summary of some of the most recent papers is found in
table \ref{tab:Parameters}.
The data examined in this paper span a baseline of about 13 years,
from 1998 to 2011. In the following sections we will try to 
explain the long-term spin evolution with a self consistent model
  that can explain the measured strength of the magnetic field, the
  neutron star spin frequency derivative and the mass transfer
  rate. Then we will move on to comment on the different possible
  magnetic fields reported by \cite{Iaria_2015A&A...577A..63I} and
  \cite{Sasano_2014PASJ...66...35S}. We then proceed to discuss
  whether the observed spin frequency derivative reflects a secular
  spin up that will continue in the future, or whether we are indeed
  just observing a short-term spin up that will change sign and or
  magnitude in the future.

\subsection{Long term spin evolution} \label{sec: discussion - LT spin}

The large spin up of 2A 1822-371 implies a very short spin-up time scale
of about $\nu/\dot{\nu}\approx\,7000$yr. This is an extremely short
time scale for a system that should take several million
years to spin up \citep{Bhattacharya_1991PhR...203....1B}.
The relatively constant spin up
implies no variation in the accretion torques acting upon the
neutron star at least down to timescales of 8-10 days.
However, as
shown in Section \ref{sec:LT_spin} the linear fit to the spin
frequency vs. time gives a very large $\chi^2$ which implies that the
spin frequency is not increasing linearly with time. This would be
expected in case of constant accretion torque, but the observed
flux has an rms fluctuation of the order of 4.4 mCrab,
suggesting that accretion torques do indeed vary slightly with time.
The strongest evidence for this comes from the last data point in 2011 where
the spin up increase by a factor 3. Accretion theory predicts that the
strenght of the spin up should scale with the amount of mass accreted
according to the following relation (see e.g., Bildsten et al. 1997):
\begin{equation}
\dot{\nu}\propto \dot{M}^{6/7}.
  \end{equation}

If the mass accretion rate is related to the X-ray luminosity (and
thus the X-ray flux) with the usual relation $L_X \approx
\eta\,c^2\dot{M}$ then we should expect
$\dot{\nu}\propto\,F_X^{6/7}$. Therefore a 3$\times$ larger
$\dot{\nu}$ should imply a $\approx3\times$ larger X-ray
flux. However, the average 2--16 keV X-ray flux is not varying, on
average, by this amount and therefore this probably means that the
2--16 keV X-ray flux is not a good tracer of the instantaneous mass
accretion rate.  Such possibility has been proposed to explain the
behaviour of a number of other LMXBs \citep{Klis_2001ApJ...561..943V}.

\begin{table*}
	\centering
	\caption{The different parameters in the most recent published papers on 2A 1822371. The papers are JvdK2001 = \citet{Jonker_2001ApJ...553L..43J}, Jain2010=\citet{Jain_2010MNRAS.409..755J},BU2010=\citet{Burderi_2010A&A...515A..44B}, Bay2010=\citet{Bayless_2010ApJ...709..251B}, Ia2011= =\citet{Iaria_2011A&A...534A..85I}, Som2012=\citet{Somero2012A&A...539A.111S}, Sas2014=\citet{Sasano_2014PASJ...66...35S} \& Ia2015=\citet{Iaria_2015A&A...577A..63I} }
	\label{tab:Parameters}
	\begin{tabular}{lcccccr} 
		\hline
		Parameter & JvdK2001 & Jain2010 & Bay2010(UV/Optical) & Sas2014 & Ia2015 \\
		\hline
		$\dot{\nu}$ (Hz/s) & (8.1$\pm$0.1)$\times$10$^{-12}$ & (7.06$\pm$0.01)$\times$10$^{-12}$ & - & (6.9$\pm$0.1)$\times$10$^{-12}$ & (7.25$\pm$0.08)$\times$10$^{-12}$ \\
		$\dot{P}$$_{Spin}$ (s/s)   & -2.85(4)$\times$10$^{-12}$ & -2.481(4)$\times$10$^{-12}$ & - & -2.43(5)$\times$10$^{-12}$ &-2.55(3)$\times$10$^{-12}$ \\
		L$_X$ (erg/s)  & 10$^{36}$-10$^{38}$ & (2.38-2.96)$\times$10$^{38}$ & $\sim$10$^{37}$ & $\sim$10$^{37}$ & 1.26$\times$10$^{38}$ \\
		$\dot{M}$(M$_{\odot}$/yr)  & - & (4.2-5.2)$\times$10$^{-8}$ & 6.4$\times$10$^{-8}$  & - & - \\
		M$_{NS}$(M${_\odot})$ & 1.4 & 1.4 & 1.35 & 1.4 & 1.61-2.32 \\
 	    B (G)  & 10$^{8}$-10$^{16}$ & (1-3)$\times$10$^{8}$ & - & 2.8$\times$10$^{12}$ & 8.8(3)$\times$10$^{10}$ \\
 	    Pulse Amp  & 0.25-3\% & - & - & $\sim \pm$5\% & $\sim$0.75\%\\
 	    $P_{orb}/\dot{P}_{orb}$ (yr)  & - & (4.9$\pm$1.1)$\times$10$^{6}$ & (3.0$\pm$0.3)$\times$10$^{6}$  & - & -\\
 	    $P_{s}$ (s) & 0.59325(2) & 0.5926852(21) & - & 0.592437(1) & 0.5928850(6) \\
 	    Time span & 1996-1998 & 1998-2007 & 1979-2006  & 2006 & 1996-2006 \\
		\hline
	\end{tabular}
\end{table*}

We now wish to test if we can find a self consistent model that
explains the observed spin parameters of 2A 1822-371.
First, let's define the
co-rotation radius as that point in the accretion disk where matter
has the same angular velocity of the neutron star:
\begin{equation}\label{eq:Rco}
R_{CO}=\left(\frac{GM_{NS}}{4\pi^2 \nu^2}\right)^{1/3}
\end{equation}

Then, following \citet{Ghosh_1979ApJ...234..296G}, we can define the
magnetospheric radius as the point where the ram pressure of the disk
plasma equals the magnetic pressure:
\begin{equation}\label{eq:Rm}
R_m=\xi R_A=\xi\left(\frac{\mu^4}{2GM_{NS}\dot{M}^2}\right)^{1/7}
\end{equation}
Such definition has some issues since it is derived by equating the
dipolar magnetic field pressure to the ram pressure of a spherically
symmetric free falling gas. The factor $\xi\approx 0.5$ is typically
used to account for the disk geometry instead of the spherical
symmetry of the free falling gas. Several works
(\citealt{Aly_1985A&A...143...19A, Lovelace_1995MNRAS.275..244L},
\citealt{Goodson_1997ApJ...489..199G}) have demonstrated that the
magnetospheric-disk interaction will quickly generate an azimuthal
field component that causes many field lines to open and reconnect at
infinity. \citet{Dangelo_2010MNRAS.406.1208D,Dangelo_2012MNRAS.420..416D}
have thus derived a different version of the magnetospheric radius
which accounts for this differences. The magnetospheric radius thus
obtained however, is not substantially different from the expression
above and since we are giving only an order of magnitude estimate of
the quantities we will continue to use the definition above. This
makes also the comparison with other works more direct, since they
mostly rely on the definition of magnetospheric radius given by
\citet{Ghosh_1979ApJ...234..296G}

From the table \ref{tab:Parameters}, we can see that a few parameters
reported in the literature (e.g., B, L$_x$, $\rm \dot{P}_s$, etc.) are
not consistent with each other and sometimes some reported values are
even inconsistent from a physical point of view.  For example,
magnetic field strengths as large as $10^{16}\rm \,G$ have been
discussed by \citet{Jonker_2001ApJ...553L..43J}, by using the relation between B and $\mu$, which, can only be used when $R_{m}<R_{CO}$,
since there will be no accretion if the magnetospheric radius is
outside of the co-rotation radius (with the exception of accretion
induced by magnetic diffusivity, see
e.g.,\citet{Ustyugova_2006ApJ...646..304U}). For 2A 1822-371, the
co-rotation radius is at $\approx 1200$ km, so that any
magnetospheric radius larger than this value cannot be inferred by
using Eq. \ref{eq:Rco}. With a B field of $10^{16}\rm G$ one would
indeed obtain a magnetospheric radius of $10^6$ km for a luminosity of
$10^{36}\rm erg\,s^{-1}$.

 Since we know $\dot{\nu}$ we can use
the relation for the X-ray luminosity to find the mass accretion rate
($\dot{M}$) \citep{Frank_2002apa..book.....F}:
\begin{equation}\label{eq:Mdot}
\dot{M}=\frac{L_X R_{NS}}{GM_{NS}}
\end{equation}
This assumes that the X-ray luminosity does trace the instantaneous mass accretion rate, 
which is, however, not the case in 2A 1822-371 as we have shown above.
To obtain an expression for $\mu$ that we have used in Eq. \ref{eq:Rm}
we use the angular acceleration as a function of $\dot{M}$,
$\dot{\Omega}=2\pi \dot{\nu}=\frac{\dot{M}\sqrt{GM_{NS}R_m}}{I}$
which gives:
\begin{equation}\label{eq:nudot}
\dot{\nu}\approx\,4.1\times10^{-5}\dot{M}\,M_{NS}^{1/2}\,R_m^{1/2}\,I^{-1}\mbox{Hz/s}.
\end{equation}

In the above equation we use the measured long-term $\dot{\nu}$,
and we assume a neutron star mass of
1.4M$_{\odot}$, a neutron star radius of 10 km, the moment of inertia
($I=10^{45}\rm\,g\,cm^2$, $\xi$=0.5. We assume 1.4M$_\odot$ in the above, even though \citet{Iaria_2015A&A...577A..63I} find the mass of the neutron star to be (1.69$\pm$0.13)M$_\odot$ and find the companion mass to be (0.46$\pm$0.02)M$_\odot$, within the limits given by \citet{Munos-Darias_2005ApJ...635..502M}.

We can see that the models that satisfy the condition $R_{m}<R_{CO}$
are those where the luminosity is in excess of the Eddington limit.
However, it is not known if the star really does accrete at near the
Eddington limit. The observed X-ray luminosity is only 
L$_X$$\approx$10$^{36}(d_2)^{2}$erg s$^{-1}$ assuming a (poorly
constrained) distance of 2 kpc. The best distance approximation is
between 1-5 kpc \citep{Mason_1982ApJ...262..253M,
  Parmar_2000A&A...356..175P} which
however, would shift the luminosity by less than an order of
magnitude. The most compelling evidence that the X-ray luminosity is
indeed higher than the observed value is that the ratio
L$_x$/L$_{opt}\approx$15-65 and not 500-1000 as observed in other
LMXBs \citep{Griffiths_1978Natur.276..247G,
  Bayless_2010ApJ...709..251B, Iaria_2015A&A...577A..63I,
  Somero2012A&A...539A.111S}. This means that either the optical
luminosity is much larger than expected or that only a small ($1-10\%$)
of the total X-ray luminosity of the source is effectively observed.

A source accreting at the Eddington rate requires a mass accretion
rate of $\dot{\rm M}\sim$10$^{-8}$M$_\odot$/yr. This is 2 orders of
magnitude larger than what is expected from binary evolution models
under the assumption that the donor is a main sequence star that
started Roche lobe overflow with a mass $\lesssim 1\msun$ and that the
binary evolution is driven by angular momentum loss via magnetic
braking~\citep{Podsiadlowski_2002ApJ...565.1107P}. Since 2A 1822-371
is a persistent source, the mass accretion rate (from the disk to the
neutron star) must be equal (or very close to) the mass transfer rate
(from the donor to the accretion disk).  Therefore in this model the
system will only survive for about 1 Myr.

It is insteresting to compare the behaviour of \src\, with that of
Terzan 5 X-2, another pulsar that is in many ways similar to 2A
1822-371. Terzan 5 X-2 is an 11 Hz accreting pulsar which is
accreting from a sub-giant companion ($M\gtrsim\,0.4\,M_{\odot}$) in a
relatively large orbit (orbital period of 21 hr). The neutron star has
a dipolar magnetic moment in the range of $\mu
\simeq$10$^{27}$-10$^{28}$Gcm$^3$ \citep{Cavecchi_2011ApJ...740L...8C,
  Papitto_2011A&A...526L...3P, Patruno_2012ApJ...752...33P}. This
system has a clear spin-up of $\dot{\nu}\sim$10$^{-12}$Hz s$^{-1}$,
and it appears to be evolving towards a millisecond pulsar in a very
short timecale of a few tens of million years
\citep{Patruno_2012ApJ...752...33P}. Both 2A 1822-371 and Terzan 5
X-2 seem to be at odd with the very long phases that binaries with
similar spin and orbital parameters spend in Roche lobe contact, which
can last for about 1 Gyr or more.  \citep{Patruno_2012ApJ...752...33P}
proposed that Terzan 5 X-2 is in an exceptionally early RLOF phase
although the reason why we are witnessing this unlikely event
remains an open problem.  Indeed observing two pulsars being recycled
in an early RLOF phase in the relatively small population of LMXBs is
unlikely. This means that the exceptionality of Terzan 5 X-2 cannot be
due to chance alone and there must be a common evolutionary process
that creates this kind of accreting pulsars. By using the proper
motion, the radial velocity and the current position of 2A 1822-371,
it is possible to find the original position and give an estimate of
the age of the system. \citet{Maccarone_2014MNRAS.440.1626M} found 2A
1822-371 to likely originate from close to the Galactic center, and
reported an age of about 3--4 Myr, which indeed makes the system very
young \citep{Maccarone_2014MNRAS.440.1626M} (although there is a quite
big uncertainty due to the poorly constrained distance of the
system). This may support the analogy that the system is similar to
Terzan 5 X-2 and they both are in an early Roche Lobe overflow phase.

\subsection{Torque reversal} \label{sec: discussion-torqe teversal}

2A 1822-371 share a few common features with other accreting pulsars,
besides Terzan 5 X-2. Short spin-up time scales are seen also in the
ultra-compact binary 4U 1626-67
\citep{Chakrabarty_1997ApJ...474..414C} which has $\nu /
\dot{\nu}$=5,000 yr
\citep{Chakrabarty_1997ApJ...474..414C,Beri_2014MNRAS.439.1940B}. 4U
1626-67 is a quite different binary from 2A 1822-371 since it is an
ultra compact system ($P_{\rm\,orb}\approx\,42$ min), the companion star
is not a main sequence star, but rather a degenerate He or CO white
dwarf, and the neutron star has a spin of 7.66 s. The system was
originally discovered in 1972 by \cite{Giacconi_1972ApJ...178..281G}.
Torque reversal of the system was observed for the first time in 1990,
where the system was found to be spinning down rather than up, as previously observed
\citep{Chakrabarty_1997ApJ...474..414C, Beri_2014MNRAS.439.1940B}. %

The torque reversal phenomenon is not very well understood since it is
unclear what its origin is and what triggers it. However, the
accretion torque in 4U 1626-67, no matter the sign of the spin
frequency derivative, is very steady on timescales of years, which
means that the accretion is almost certainly from an accretion
disk. \cite{Chakrabarty_1997ApJ...474..414C} even found that the
pulsar seems to be accreting steadily during spin-down. The first
phase of spin-up lasted for at least 13 yr, the spin-down then lasted
about 18 yr, and in 2008 4U 1626-67 began spinning up again
\citep{Beri_2014MNRAS.439.1940B}. In 4U 1626-67 the torque reversals
are accompanied by sudden variations in the X-ray luminosity
\citep{Chakrabarty_1997ApJ...474..414C}. A decrease in X-ray flux was
seen when the neutron star moved from a spin-up to spin-down phase,
and again there was an increase by a factor 2 in the X-ray flux with
the second torque reversal \citep{Beri_2014MNRAS.439.1940B}.  The very
short spin-up timescale found for 2A 1822-371 is thus not incompatible with
the notion that this system too will show a torque reversal
somewhere in the near future.

\subsection{Accretion Disc Corona}\label{sec:discussion-ADC}

From the spectral analysis of 2A 1822-371 and from the shape of the
eclipses it is evident that there is some extended X-ray emitting region
around the pulsar, often assumed to be the ADC, that scatters the
light originating on the pulsar \citep{White_1982ApJ...257..318W,
  Heinz_2001MNRAS.320..249H, Iaria_2013A&A...549A..33I}.

One possible scenario is that an extended Comptonizing region is in fact the
accretion curtain as it falls towards the neutron star, and it is possible
that the emission we see is actually produced by photons upscattered
through this curtain. To investigate this more quantitatively, we
build a toy model for the spectrum, in which the underlying emission
from the star (which we assume to consist of a blackbody and
Comptonized hard X-ray component) is upscattered by hot electrons in
the infalling accretion curtain. We use the method and code of
\cite{D'Angelo_2008A&A...488..441D} to produce the final Compton
upscattered spectrum. We start with an initial input spectrum (assumed
originating from the stellar surface) of a blackbody plus an
additional power-law component with a cutoff at high energies. We then
use this as a seed photon spectrum to generate an output spectrum as a
result of inverse Compton upscattering through a hot (e.g. $\gtrsim
10$keV) thermal electron cloud.

To model this process we use a Monte Carlo Compton scattering code,
whose details are described in \cite{Giannios_2004A&A...427..251G}. Briefly,
the code works by using the seed photon spectrum as the initial photon
energy distribution, and then calculating the outcome (final photon
energy and direction) of a seed photon inverse Compton scattering off
a hot electron. The electron cloud energy distribution is assumed to
be thermal, and the temperature of the cloud is an input parameter of
the simulation. The cloud is assumed to be isotropically surrounding
the source of seed photons, and the probability of scattering
depends mainly on the optical depth of the electron cloud (another
input parameter).

The model thus has six free parameters, four for the input spectrum:
the blackbody temperature ($t_{\rm bb}$), the power-law slope
($\Gamma$) and cut-off energy ($E_c$), the relative strength of the
blackbody to power law ($N$), and two for the Comptonizing cloud: its
temperature ($T_e$) and optical depth ($\tau$). We vary these
parameters in order to explore the range of temperatures and optical
depths for the electron cloud that could be made consistent with the
observed X-ray spectrum.

The requirement of an ADC has been introduced in the literature to
explain the excess of light seen during the eclipses, with the X-ray
flux never reaching a value of zero as expected from a full
eclipse. The X-ray flux is seen hovering at around 50\% of its
non-eclipse value. Furthermore the very long duration of the partial
eclipses (about 80\% of the orbit) requires an extended source
surrounding the central X-ray source \citep{White_1982ApJ...257..318W,
  Hellier_1990MNRAS.244P..39H}. The ADC was suggested to be formed
from evaporated material in the accretion disk
\citep{White_1982ApJ...257..318W}. \cite{Shakura_1973A&A....24..337S}
suggested that the central source of X-ray binaries could evaporate
material from the disk, and that if the material does not escape the
system, it could form a corona-like cloud around the central
source. \cite{White_1982ApJ...257..318W} showed that the central
source is always obscured if the inclination angle of the system is
more than 60$^\circ$, which is compatible with what is observed for 2A
1822-371, with an inclination angle of $i=$82.5$^\circ
\pm$1.5$^\circ$ \citep{Heinz_2001MNRAS.320..249H}. As we have shown in
section \ref{sec:FA} this scenario is compatible with the behaviour of
the pulsed fraction as a function of the orbital phase. Indeed the
fractional amplitude of the pulsations is consistent with being
constant regardless of the orbital phase of the binary. This implies
that the neutron star is obscured by some material throughout the
orbit. Another important point is that the fractional amplitude of the
pulsar does not vary with the depth of the eclipses.  This does
support the idea that the surface of the neutron star is never
observed and that the pulsations we do see are indeed scattered
through some medium, e.g., an ADC or an accretion stream.

\begin{figure}
\centering
\includegraphics[width=1.0\linewidth]{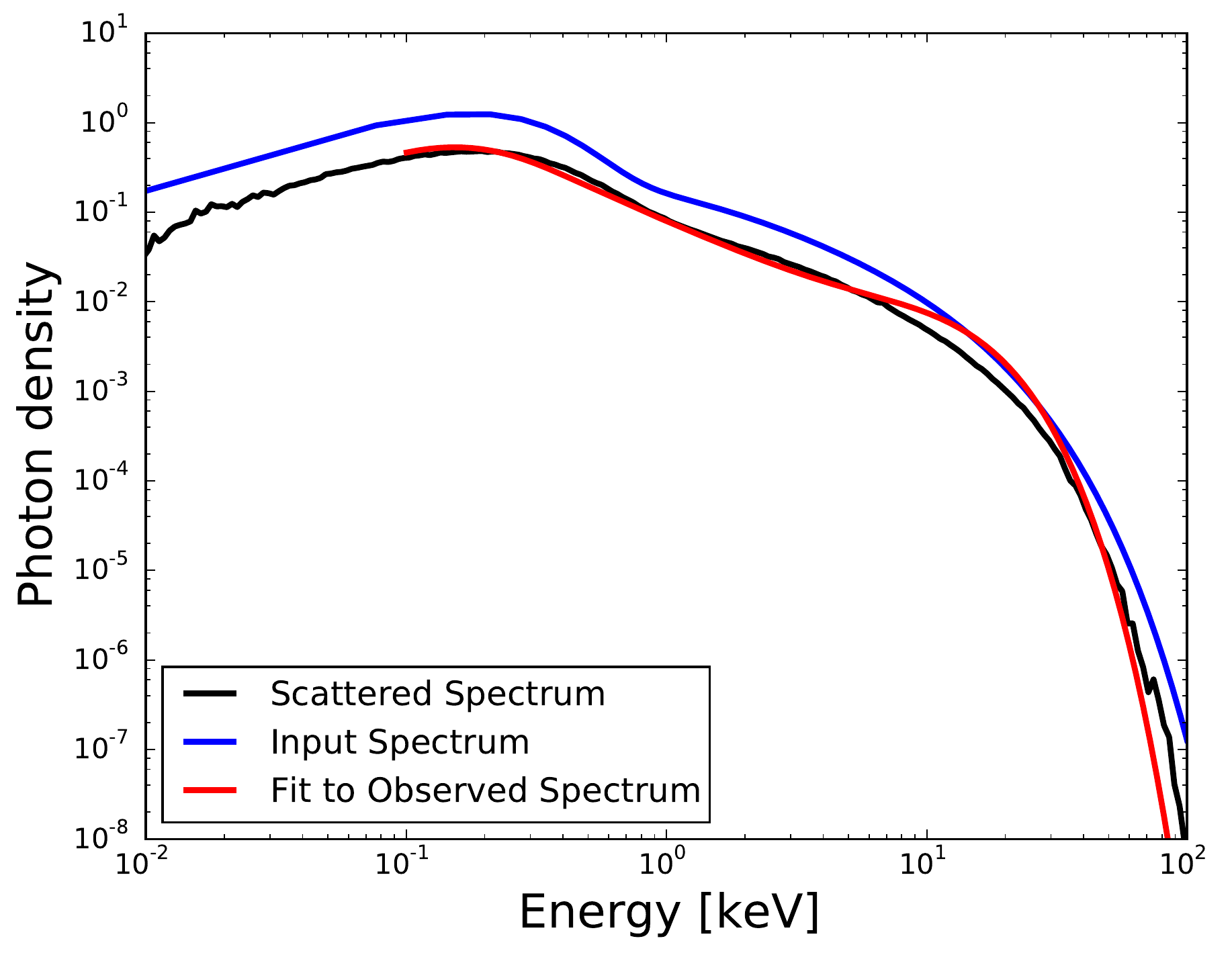}
\caption{The blue line represents the best simulated spectrum. The blue line is our guess corresponding to the red line, which is the fit to the spectrum of X 1822-371 used by \citet{Iaria_2015A&A...577A..63I}. The green line is the input spectrum we have used, consisting of both a black body and a power law component.} 
\label{fig:Spectrum_tau1}
\end{figure}

\citet{Iaria_2015A&A...577A..63I,Iaria_2001ApJ...557...24I} and
\citet{Parmar_2000A&A...356..175P} previously suggested that the ADC
must be optically thick. This was also discussed by
\citet{Heinz_2001MNRAS.320..249H}, and
\citet{Iaria_2013A&A...549A..33I}, who stated that an optically thick
ADC is not consistent with the pulsations observed. To explore the possibilities for a lower value for the
optical depth we explored the parameter space for the
optical depth, the power law index and the electron temperature of the
Compton up-scattering cloud. We found it possible to create an input
spectrum, that, sent through the Compton up-scattering cloud, would be
similar to the fitted spectrum with a high optical depth used by
\cite{Iaria_2015A&A...577A..63I}. This can be seen on
Fig. \ref{fig:Spectrum_tau1}, where the red line is the fit used by
\cite{Iaria_2015A&A...577A..63I}, the blue line is the input spectrum
we created, using both a power law and a black body, and the black
line is the output spectrum after the input spectrum has been sent
through a Compton cloud. 

The blue line on Fig. \ref{fig:Spectrum_tau1} corresponds to a power law
index of $\Gamma=1$, electron temperature of $E_c=10$keV and 
optical depth of $\tau=1$. Our test of the optical depth should be
taken only as a proof of principle, that a model spectra with a black
body and a power law can recreate the spectra observed \textit{even if the
  optical depth is small}. The quality of the
spectra and its match with the spectra found for example by
\cite{Iaria_2015A&A...577A..63I} is judged by eye and by the normalized root mean square deviation which we find to be 28\%.

The limits for the electron temperature is $3<T_e<15$ keV, the power
law index limits are $0.5<\Gamma<1.5$ and the acceptable range of
optical depth is $0.01<\tau<3$. Within these limits the spectra are
reasonably reproduced with a much lower optical depth that is thus
compatible with the presence of pulsation without requirement an
optically thick Comptonization region plus an optically thick
scattering ADC as suggested in \citet{Iaria_2013A&A...549A..33I}.  These results are
also compatible with recent spectral modelling performed by \citet{Niu_2016RAA....16...57N} which discussed the possibility that the accretion
disk corona around 2A 1822-371 is indeed optically thin.

The fit we created give a simple structure of the system, where the
ADC is surrounding the pulsar, and reaching further out than the
secondary star, making the pulsar light coming from an extended
source, and thus both explaining the pulsations and the 50\% depth of
the eclipses.  The structure of the system we suggest is initially
very simple when compared to what \cite{White_1982ApJ...257..318W}
suggested. By analysing a spectrum that consists of both a power law
and black body component, it is possible to eliminate the optically
thick Comptonizing region, and thus explain the system only with an
optically thin (or moderately thick; $\tau\approx 0.01-3$) ADC
surrounding the pulsar. Another possibility for the geometry is that
we could be seeing this system through an extended accretion
stream. This stream scatters the light and pulsations just as an ADC
would do. Assuming this as an explanation, simplifies the structure of
the system even further.\\

2A 1822-371 is not the only suggested ADC systems. Another such
system is MS 1603.3+2600. MS 1603.3+2600 was discovered by
\cite{Morris_1990ApJ...365..686M} and has a 1.7 hr orbital period
\citep{Jonker_MS1603_2003MNRAS.346..684J}. The source shows both
similarities and differences with 2A 1822-371. Both sources appear to
be at a fairly high inclination, although the precise value for the
inclination angle is not known for MS 1603.3+2600. MS 1603.3+2600 is
thought to be at a distance of about 6--24 kpc
\citep{Jonker_MS1603_2003MNRAS.346..684J, Hakala_2005MNRAS.356.1133H,
  Parmar_2000A&A...356..175P}. Despite not showing accretion powered
pulsations, MS 1603.3+2600 has a neutron star primary, since the
source shows type I X-ray bursts
\citep{Jonker_MS1603_2003MNRAS.346..684J, Hakala_2005MNRAS.356.1133H}.
The existence of an ADC around the source is supported by variations
in the X-ray bursts. \citet{Hakala_2005MNRAS.356.1133H} looked at two
XMM-\textit{Newton} observations taken on 2003 January 20 and 22. They
observed several Type I X-ray burst candidates. The bursts during the
first observation have a count rate of about 6 counts s$^{-1}$ whereas
those seen during the second part of the observation only have a count
rate of 2-3 counts s$^{-1}$. The variation suggests that the bursts
are not directly observed, thus it is possibly only scattered X-rays
that are observed, with the scattering medium forming an ADC
\citep{Hakala_2005MNRAS.356.1133H}.

\subsection{2A 1822-371 as a super Eddington source}

A big weakness of the model that we have discussed so far is that it
requires an Eddington limited accretion rate, despite the binary
containing a low mass main sequence star that should transfer mass at
a rate of about $10^{-10}\rm\,\msun\,yr^{-1}$.
\citet{Jonker_2003MNRAS.339..663J} performed detailed optical
observations of \src\, and suggested that one possible interpretation
of the spectroscopic results is that the donor star is out of thermal
equilibrium. \citet{Cowley_2003AJ....125.2163C} also suggested that
the donor must be somewhat evolved since otherwise it would not fill
its Roche lobe. \citet{Munos-Darias_2005ApJ...635..502M} also proposed
that 2A 1822-371 is a LMXB which descends from an intermediate mass
X-ray binary progenitor (initial $M\gtrsim\,1\,M_{\odot}$), with the
companion that has already lost a substantial amount of mass. In this
case the mass transfer would not be driven by angular momentum loss
via magnetic braking and/or gravitational radiation but would proceed
on the thermal (Kelvin-Helmotz, KH) timescale of the companion (see
e.g., \citealt{King_1996ApJ...467..761K}).

The thermal timescale is $\tau_{KH}\approx\,GM_2^2/(2R\,L_{nucl})$ where
$G$ is the universal gravitational constant, $M_2$ is the companion
mass, $R_2$ its radius and $L_{nucl}$ the nuclear stellar luminosity. The stellar
luminosity is not well known in \src\, since the optical observations
are usually dominated by the disk emission/irradiation. Furthermore, as shown by
\citet{King_1995ApJ...444L..37K}, the thermal timescale changes when 
irradiation is present, since the stellar surface luminosity might exceed
the nuclear luminosity. Since the observed X-ray luminosity is of the
order of $10^{36}\ergs$ and assuming that all the irradiation luminosity
is re-emitted by the stellar surface of the donor, the irradiation luminosity
would, due to geometric effects, correspond to approximately $10^{34}\ergs$ and the KH timescale becomes
$\tau_{KH}\gtrsim\,10^7$ yr, using M$_2$=0.46M$_\odot$ and R$_2$ corresponding to the Roche Lobe radius, R$_L$=0.6R$_\odot$.
If the companion star evolved on the $\tau_{KH}$ timescale, then the mass transfer rate can be as high as ${\sim}10^{-7}\rm\,M_{\odot}\,yr^{-1}$

We now try to interpret the observations of \src\, in light of this
hypothesis. In the following there is a big uncertainty on most parameters, thus the results could vary within an order of magnitude, and are only approximate.
The orbital separation of \src\, changes for two reasons:
1. the redistribution of angular momentum in the binary and 2. the
loss of angular momentum via magnetic braking/gravitational wave
emission (see e.g., \citet{Frank_2002apa..book.....F}):
\begin{equation}\label{eq:adot}
\frac{\dot{a}}{a} = \frac{2\dot{J}}{J}+\frac{-2\dot{M}_{\rm 2}}{M_{\rm 2}}\left(1-q\right)
\end{equation}
here $q=M_{\rm 2}/M_{\rm NS}$ is the mass ratio between the companion
($M_{\rm 2}$) and neutron star mass ($M_{\rm NS}$), $a$ is the orbital
separation, $J$ is the angular momentum and the dot refers to the
first time derivative. In J we include all effects, e.g. magnetic breaking, gravitational waves and mass loss \citep{Tauris_2006csxs.book..623T}. If the mass transfer $\dot{M}_{\rm
  2}\approx\,10^{-7}\,M_{\odot}\,yr^{-1}$ then, assuming to first
order a conservative mass transfer scenario ($\dot{J}\approx0$) we
expect a relative variation of the orbit $\dot{a}/{a}\sim
10^{-14}\,s^{-1}$. The observations of the orbital period derivative
$\dot{P}_{\rm orb}\approx 1.51\times10^{-10}$ \citep{Iaria_2011A&A...534A..85I} provide a
direct test of this hypothesis. Indeed from the 3-rd Kepler law we
expect that $\dot{a}/{a} = 2\dot{P}_{\rm orb}/3P_{\rm orb}$ and the
observed values give: $2\dot{P}_{\rm orb}/3P_{\rm orb}\approx5\times
10^{-15}\rm\,s^{-1}$ which is in good agreement with the hypothesis
that \src\, is evolving on a thermal timescale in a conservative mass
transfer scenario.

Since the mass transfer is super-Eddington we can follow the line of
reasoning of \citet{King_2016MNRAS.458L..10K}, where they examine the
case of the Ultra-Luminous X-ray source (ULX) M82 X-1. If that donor
in M82 X-1 is in a super-Eddington mass transfer phase then the mass
accretion rate, magnetic radius and the magnetic moment of the neutron
star ($\mu$) can be inferred from first principles.  We argue here
that one remarkable possibility to explain the phenomenology of \src\,
is that it is a mildly super-Eddington source.

In this case the accretion disk will be the standard Shakura-Sunyaev
geometrically thin disk down to the so-called spherization radius
\citep{Shakura_1973A&A....24..337S}:
\begin{equation}\label{eq:sph}
  R_{\rm sph}= \frac{27}{4}\frac{\dot{M}_{\rm tr}}{\dot{M}_{\rm Edd}}R_{\rm g}
\end{equation}
where $\dot{M}_{\rm tr}$ and $\dot{M}_{\rm Edd}$ are the mass transfer
rate ($10^{-7}\rm\,M_{\odot}\,yr^{-1}$) and the Eddington limit for a
neutron star (that we set equal to $\approx2\times
10^{-8}\rm\,M_{\odot}\,yr^{-1}$) and the $R_{\rm
  g}=GM/c^2\approx2\times10^5\rm\,cm$ is the neutron star gravitational radius ($M_{1}=1.69\rm\,M_{\odot}$).  This gives a
spherization radius of $\sim10^7$ cm. Beyond this radius the
flow will become geometrically thick and generate a funnel flow that can produce beaming (see e.g., \citet{King_2009MNRAS.393L..41K}).

\begin{figure}
\centering
\includegraphics[width=1.0\columnwidth]{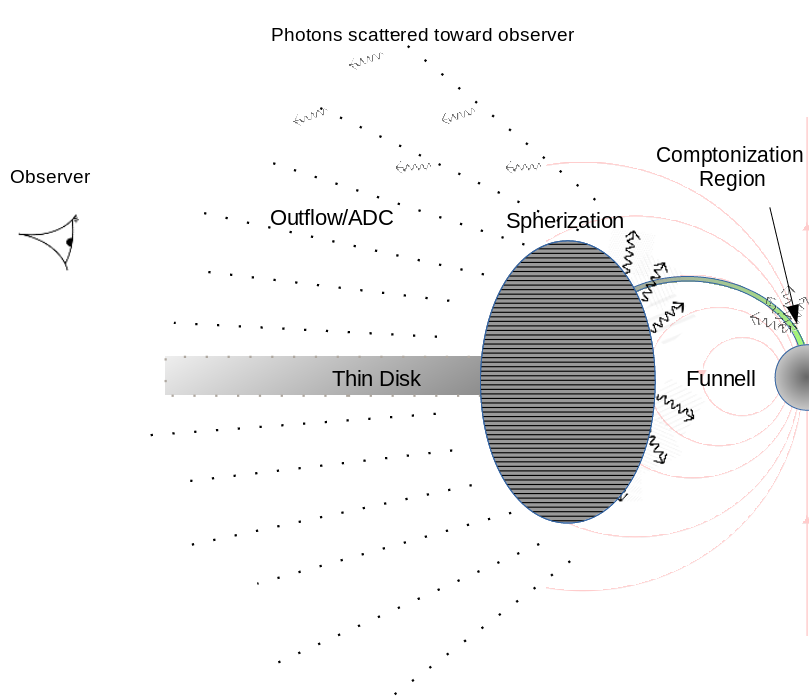}
\caption{The illustrations show the geometry of the problem according to the super-Eddington scenario that we propose for \src\,. The observer is located at an inclination angle $i=82^{\circ}$ and we have chosen an aligned rotator for simplicity. The thick part of the accretion disk begins at the spherization radius and ends at the magnetospheric radius where the plasma becomes channeled towards the neutron star poles. The accretion flow along the magnetic field lines is abruptely stopped a the neutron star surface and a shock forms close to the neutron star surface. It is in this shock that the Comptonization process takes place. The outflow generates instead an optically thin cloud around the neutron star that is responsible for the scattering of a small portion of the X-ray photons towards the direction of the observer.}
\label{fig:illustration}
\end{figure}

If the innermost region of the accretion disk is truncated at the
magnetospheric radius $R_{\rm M}$, with $R_{\rm M}<R_{\rm sph}$, then the local mass accretion rate in any annulus of the disc with radius $R<R_{\rm sph}$ needs to be:
\begin{equation}\label{eq:mdot}
  \dot{M}(R)=R/R_{\rm M}\dot{M}_{\rm Edd}.
\end{equation}
Together with the expression for the magnetic dipole moment and spin
up (Eq.~\ref{eq:nudot}), the spherization radius and the
mass accretion rate equations (Eq.~\ref{eq:sph}, \ref{eq:mdot}) form a
set of 4 equations with four variables, $R_{\rm sph}$, $\dot{M}(R_{\rm
  M})$, $\mu$ and $R_{\rm M}$. Following the method by \cite{King_2016MNRAS.458L..10K} one finds $R_{\rm M}\sim
10^6-10^7$ cm, $\mu\approx 1\times10^{28}\rm\,G\,cm^3$
(corresponding to a magnetic field of $\approx2\times10^{10}\rm\,G$ at the
poles and assuming a $\dot{M}(R_{\rm M})\approx\,2\times\,10^{-8}\rm\,M_{\odot}\,yr^{-1}$).
The fact that the spherization radius and the magnetospheric radius
are very close suggests that only the innermost portions of the
accretion disk must be geometrically thick and generate a strong outflow. Such outflow might be
responsible for the observed ADC since it will surround the central
X-ray source.  The value of the magnetic field is somewhat
smaller than the one inferred from the possible cyclotron line reported
by~\citet{Iaria_2015A&A...577A..63I} although it is of the same order of magnitude despite the large uncertainties involved. Indeed we suggest that:
\begin{itemize}
\item The donor star of \src\, is irradiated by a luminosity of $\approx 10^{36}\ergs$.
\item The irradiation drives a thermal timescale mass transfer of the order of $10^{-7}\mdot$.
\item The super-Eddington mass transfer rate generates an outflow at the spherization radius, very close to the magnetospheric radius.
\item The inner regions of the disk are geometrically thick and obscure the central source, as seen in Fig. \ref{fig:illustration}. 
\end{itemize}

Since the donor (and the observer) are nearly parallel to the
accretion disk plane, even a very mild beaming will be sufficient to
direct most of the radiation outside the line of sight of both the
donor and the observer. Therefore the donor will not be irradiated by
an X-ray luminosity much larger than the observed $L_{\rm X}\approx
10^{36}\ergs$.
 
Finally, as a self-consistency check we need to verify whether the
angular momentum loss due to the expulsion of material from the Roche
lobe of the neutron star can alter significantly the orbit of the
binary. As a limiting case we assume that all the material transferred
in the neutron star Roche lobe is expelled and thus, following \citet{Postnov_2006LRR.....9....6P}:
\begin{equation}
\frac{\dot{J}_{\rm orb}}{J_{\rm orb}} =\beta \frac{\dot{M}_2\,M_2}{M_{\rm NS}M_{\rm T}}\approx 10^{-16}\rm\,s^{-1}
\end{equation}
where $M_{\rm T}=M_2 + M_{\rm NS}$ is the total binary mass (assumed
here to be approximately 2$\msun$) and $J_{\rm orb}$ and $\dot{J}_{\rm
  orb}$ are the total orbital angular momentum of the binary and its
variation, and $\beta$ is the fraction of the mass that is expelled. By inserting this value in Eq.~\ref{eq:adot} and by
considering that $\dot{M_2}\approx 10^{-7}\msun\rm\,yr^{-1}$ we see that the
final value of $\dot{a}/{a}$ is still of the order of
$10^{-14}\rm\,s^{-1}$ which, again, is compatible with the
observations.

\subsubsection{Possible Tests of the Proposed Model}

If \src\, is really a mildly super-Eddington source then:

\begin{itemize}
\item The neutron star magnetic field must be of the order of a few times $10^{10}$ G.
\item The source should not show any torque reversal in the near future.
\item An outflow from the neutron star must be present and possible radio emission from a jet/outflow should be expected.  
\end{itemize}

\section{Conclusion}\label{sec:conclusion}
We examined 13 years of data from \textit{RXTE}, and conclude that the
long-term spin frequency derivative and the two phase connected data
sets, where we found short-term spin frequency derivatives, support an
overall fast spin-up. The spin-up supports previous work by
\cite{Iaria_2015A&A...577A..63I, Jain_2010MNRAS.409..755J} and
\citet{Chou_2016arXiv160804190C}. We tested if there was any
flux-phase correlation present in this pulsar as there is in other
systems, but found that there were no correlation.

We propose that the \src\, is a relatively young binary (age of
$\sim1-10$ Myr) in which the donor is in a thermal timescale
mass transfer phase. The orbital variation observed can be explained
by the effect of the redistribution of angular momentum in the binary
with no need for a large mass outflow from the donor star. An outflow
is instead expected from the Roche lobe of the neutron star as a
consequence of the nearly Eddington mass accretion rate occurring
close to the neutron star magnetospheric radius. We propose that the
outflow generates a large scale optically thin corona with $\tau\approx\,1$
that surrounds the system. The lack of variability in the fractional amplitude
suggests however, that the central source is
always  partially obscured and thus an optically thick
region must form close to the neutron star at the approximate location
of the spherization radius. We propose that this optically thick region
generates a mild beaming as a consequence of the super-Eddington mass transfer
rate. The Eddington/super-Eddington luminosity is not seen directly since the
observer is viewing the source nearly edge on, in a way similar to what
happens in the black hole binary SS433~\citep{King_2009MNRAS.393L..41K}.

\section*{Acknowledgements}\label{Acknowledgements}
The authors would like to thank D. Giannios for use of the Numerical code. 
AP acknowledges support from an NWO Vidi Fellowship. The authors also thank the referee, Rosario Iaria, for useful comments.


\bibliographystyle{mnras} 
\bibliography{bibliography1}

\bsp	
\label{lastpage}
\end{document}